# Access to Healthcare for People with Alzheimer's Disease and Related Dementias

Saeed Saleh Namadi[1], Jie Chen[2], Deb Niemeier [1]

## Abstract

**Background:** Alzheimer's Disease and Related Dementias (ADRD) affects millions worldwide. Significant disparities exist in ADRD diagnosis and care, disproportionately impacting minority and socioeconomically vulnerable populations

**Objective:** In this study, we investigate the relationship between ADRD density and accessibility to healthcare. We identify underserved and overserved areas in Maryland based on diagnosed cases and mortality due to ADRD, focusing on geographic disparities in care.

**Methods:** 2023 Maryland ADRD patients were identified using ICD-10 codes from. Accessibility was measured using the Kernel Density Two-Step Floating Catchment Area (KD2SFCA) method. The Gini index and t-tests were used to analyze disparities between urban and rural areas. Hot Spot Analysis Getis-Ord Gi* and local bivariate relationships analysis were applied to assess spatial correlations. Principal component analysis (PCA) was applied to calculate the health risk index.

**Results:** Hospital accessibility was unevenly distributed. Mortality rates from ADRD were higher in underserved areas with fewer hospitals. Hot spot analysis shows eastern and southern Maryland have zones with high mortality per population and per ADRD patient, surrounded by similarly high-rate zones. Central Maryland shows lower death rates per patient but more hospital facilities. In eastern Maryland, higher-poverty areas are surrounded by zones with lower accessibility and higher health risk indices.

**Conclusion:** Hospital accessibility is unevenly distributed, creating major rural disparities. Underserved regions in terms of access to healthcare facilities, particularly in eastern and southern Maryland, exhibit high ADRD mortality rates despite low diagnosis rates. This suggests that many ADRD cases remain undiagnosed, underdiagnosed, or subject to delayed treatment.

[1] Department of Civil and Environmental Engineering, University of Maryland, College Park, Maryland, USA.
[2] Department of Health Policy and Management, School of Public Health, University of Maryland, College Park, Maryland, USA.

# Introduction

Alzheimer's Disease and Related Dementias (ADRD) represent a sequence of progressive conditions that significantly affect brain function, particularly in older adults. These diseases present challenges to memory, language, and overall cognitive abilities. Approximately 6.7 million adults in the United States currently live with ADRD, a number expected to more than double to 13.8 million by 2060, driven largely by aging demographics (1). Globally, Alzheimer's disease alone accounts for nearly 10 million new cases annually, ranking as the seventh leading cause of death among adults, an alarming statistic that continues to rise (2) (3). Compounding the disease burden, ADRD frequently coexists with conditions like cardiovascular disease, diabetes, hypertension, and depression, creating complex healthcare challenges for affected individuals (1). This growing public health crisis highlights the urgent need for the development of high-quality care plans that address the diverse needs of patients with ADRD across all types of communities.

Fairness in the prevalence and management of ADRD remains a critical concern, as significant disparities exist across populations (2) (4) (5) (6) (7). In the United States, women constitute about two-thirds of those diagnosed with AD/ADRD (8), and racial and ethnic minority groups face disproportionate risks; older Black Americans are twice as likely, and older Hispanic Americans 1.5 times as likely, to develop AD/ADRD compared to older White adults (9). These inequalities are compounded by evidence showing that minority groups often receive later-stage diagnoses, encounter delays in accessing primary care, and lack opportunities for coordinated care (10) (11). These differences in healthcare access raise significant concerns about their effect on mortality rates (12), particularly among individuals with ADRD.

ADRD touches millions globally, with mortality rates shaped by various risk factors and displaying significant regional heterogeneity (13). While immutable factors like age and family history elevate the risk of ADRD and associated mortality, doubling the likelihood of death, modifiable risk factors offer critical opportunities for prevention (14) (15) (16). Research indicates that addressing these modifiable risks could prevent or delay up to 40% of dementia cases, highlighting the potential for impactful interventions (17). The success of such interventions depends on access to healthcare services, which directly impacts patients' ability to receive timely and optimal diagnostics, treatment, and care.

Access to healthcare services, particularly hospitals, is a critical determinant of overall health outcomes (18) (19) (20) (21), yet variations in hospital supply and population demand, as well as challenges with transportation, contribute to disparities in accessibility (22) (23). Often, these differences in access disproportionately affect vulnerable populations, including low-income individuals, older adults, children, and people with disabilities, who often rely more heavily on healthcare services. Socioeconomic and geography (like rural-urban divides) can also create barriers to essential care (24). The transportation field has developed a number of ways to gauge the level of accessibility, ranging from simple proximity measures to advanced spatial interaction models. We use the two-step floating catchment area (2SFCA) method, which accounts for both supply-demand imbalances and cross-border healthcare-seeking behavior (21,24–29). The 2SFCA method calculates accessibility in two stages: first, by estimating provider-to-population ratios within catchment zones around healthcare facilities, and second, aggregating these ratios for demand locations (29) (30) (31). The method is both flexible and interpretable and is frequently used in studies mapping access to primary care, emergency medical services, and hospital care (25,32–35).

Existing studies document healthcare disparities among ADRD patients using the lens of race, ethnicity, and socio-demographic factors, but these factors are not place-based. Critical gaps in knowledge persist in spatially granular analyses of accessibility at localized geographic scales, such as ZIP codes or census tracts. This granularity is essential to understand hyperlocal differences in care access, particularly for ADRD populations whose mobility and care needs are uniquely shaped by cognitive and physical limitations. Even with researchers studying ADRD mortality rates and diagnostic disparities (36), there is a lack of studies comparing these mortality rates with diagnosed cases, a crucial step in pinpointing underserved and overserved areas to strengthen healthcare equity. Without this integration, it is difficult to identify regions where elevated mortality rates may reflect systemic underservice versus overservice areas. We also go beyond the determinants of dementia to extend understanding of how health and socioeconomic factors influence the spatial distribution of ADRD. Although geographic methodologies and spatial regression models have been applied in broader risk and environmental research (37), hot spot analysis has rarely been deployed to characterize risk, optimize care, or develop policy in public health. Bridging these methodological and conceptual gaps will advance spatially targeted policies, optimize resource allocation, and mitigate place-based disparities in ADRD care and outcomes (38).

This paper makes three contributions to the study of healthcare accessibility challenges and barriers for ADRD patients. First, we leverage two comprehensive datasets, the Healthcare Cost and Utilization Project (HCUP) and the American Hospital Association (AHA), to conduct a detailed ZIP code level analysis of ADRD patients' accessibility to healthcare facilities in Maryland. By identifying the spatial distribution of healthcare access and assessing equality across urban and rural ZIP codes, we can examine how socio-demographic differences contribute to disparities in healthcare accessibility. Second, addressing the gap in research that compares mortality rates with ADRD diagnosis rates, we evaluate the relationship between ADRD-related mortality and the number of diagnosed cases. By identifying areas with disproportionately high mortality rates but lower patient numbers, and vice versa, our analysis highlights underserved and overserved regions, ensuring that attention is directed to counties with significantly higher-than-average mortality rates. Finally, we advance the geographical analysis of ADRD distribution using hot spot analysis. This approach allows for an in-depth examination of how the relationships between key variables, such as accessibility, socio-economic factors, and healthcare disparities, vary across different geographic locations. Unlike traditional models that only establish general associations, this spatial analysis provides crucial insights into the geographic dimensions of healthcare inequality, offering valuable information for policymakers and researchers to develop targeted strategies for improving fairness in ADRD care.

# Methodology

## Data

We use the 2023 State Inpatient Databases (SID) (39) from the state of Maryland, along with the American Hospital Association (AHA) (40) Annual Survey, the AHA Information Technology Supplement (AHAIT), the CDC WONDER (Wide-ranging Online Data for Epidemiologic Research) (41), and the American Community Survey (ACS) (42). The SID consists of state-specific files that include all inpatient care records from participating states for each year. The AHA Annual Survey and AHAIT provide data on hospital characteristics (The measurement of the number of beds was considered when calculating demand in accessibility), the CDC WONDER offers dementia mortality data at the county level, and the ACS supplies sociodemographic information for each state. Our sample includes patients diagnosed with ADRD who reside in the state of Maryland. The ADRD diagnoses are identified using ICD-10 codes and are based on guidelines from the CMS Chronic Condition Warehouse (CCW) (43) (Centers for Medicare & Medicaid Services, 2022).

## Measure

We compute **accessibility** and the **number of ADRD patients at the ZIP code level**. The Kernel Density Two-Step Floating Catchment Area (KD2SFCA) method was used to calculate accessibility. The KD2SFCA method enhances traditional accessibility measures by incorporating distance decay effects within catchment areas (Formula (4)). This approach recognizes that the likelihood of availability to a healthcare service decreases as travel distance increases, providing a more nuanced understanding of spatial accessibility (44) (45). The KD2SFCA method calculates the supply-to-demand ratio at each demand location within a specified threshold. This process involves two steps, represented by the following mathematical models:

For each hospital (j): $D_j = \frac{S_j}{\sum_{k \in [d_{kj} \leq d_0]} P_k f(d_{kj})}$ (1)

For each ZIP code (i): $A_i = \sum_{j \in [d_{ij} \leq d_0]} D_j f(d_{ij})$ (2)

In the first step, for each hospital location j, all ZIP code locations k within a specified distance from hospital j were identified, and the supply-to-demand ratio $D_j$ within the catchment area was calculated. In formula (1), $d_{kj}$ represents the distance between ZIP code k and hospital j, while $d_0$ is the distance threshold defines the hospital's service range (15 miles in this study). The demand at ZIP code k, denoted as $P_k$, is represented by the number of ADRD patients, whereas the hospital's capacity $S_j$ is measured by the number of beds in the hospital. The impedance function f(d), defined in formula (3), accounts for the preference of individuals to visit closer hospitals, with accessibility decreasing as travel time increases. Common impedance functions include Gaussian, Inverse Power, and Exponential (46) with the Gaussian function selected due to its gradual decline, preventing a sharp drop to zero.

In the second step, for each ZIP code location i, all hospital locations j within the distance threshold $d_0$ are identified and we sum the supply-to-demand ratios $D_j$ at these hospitals to determine the accessibility $A_i$ at ZIP code i. This step also considers the attenuation of supply by applying the impedance function f(d) in formula (2). Higher values of $A_i$ indicate better accessibility, while lower values suggest areas with healthcare shortages.

$$f(d) = \begin{cases} e^{-\frac{1}{2}*\left(\frac{d}{d_0}\right)^2} - e^{-\frac{1}{2}}, d \leq d_0 \\ 0, d_{ij} > d_0 \end{cases} \quad (3)$$

$$A_i = \sum_{j \in [d_{ij} \leq d_0]} \frac{S_j f(d_{ij})}{\sum_{k \in [d_{kj} \leq d_0]} P_k f(d_{kj})} \quad (4)$$

The information on patients, including their ZIP code of residence and type of disease, was obtained from the Healthcare Cost and Utilization Project (HCUP) data. This study identified primary ADRD diagnoses based on underlying causes classified under the International Statistical Classification of Diseases and Related Health Problems, Tenth Revision (ICD-10) codes. These codes included unspecified dementia (F03), Alzheimer's disease (G30), vascular dementia (F01), and other degenerative diseases of the nervous system not otherwise classified (G31) (43)

Our socio-demographic variables are sourced from the ACS dataset (42). These variables include the percentage of each race, poverty rate, age, percentage of people with health insurance and percentage of people with higher education at the ZIP code level. Additionally, variables related to hospital characteristics were obtained from the AHA data, such as the number of beds in each hospital.

The health risk index was calculated using Principal Component Analysis (PCA).(47) PCA is a dimensionality reduction and machine learning method used to simplify a large dataset into a smaller set while preserving significant patterns and trends. (48) (49) To enable the application of the Local Bivariate Relationships method, PCA combines diseases such as diabetes, obesity, asthma, depression, hyperlipidemia, hypertension, and heart disease into a single composite measure, the health risk index.

### Analysis

**Table *1*** presents the distribution of ADRD patients in Maryland. After filtering the data using ICD-10 codes to identify ADRD patients, the KD2SFCA method was applied to 1,455 patients in the state. The Gini index was used to demonstrate equality of accessibility between urban and rural areas. A t-test analysis of the distribution of variables in urban and rural areas highlighted significant differences between these regions. Furthermore, to identify underserved and overserved areas in terms of access to healthcare facilities, mortality rate data at the county level were analyzed and compared with the distribution of ADRD diagnoses Underserved areas are considered to have limited access to outpatient care and specialized services, which may contribute to delayed detection, poorer care coordination and management, lower healthcare quality, and reduced treatment rates. Counties with significantly elevated death rates were identified. In the final step of this research, hot spot analysis along with statistical methods such as local bivariate relationships and the Getis-Ord Gi*, were applied to examine the spatial correlation of the data and to better understand the relationship between variables and the distribution of ADRD patients across Maryland.

# Results

We have a total of 384,252 patients in 2023 in the state of Maryland. Among them, 1,455 individuals were primarily diagnosed with ADRD, and we use these for our analysis. These patients incurred over $47 million in hospital inpatient charges in just the year 2023 alone. **Table 1** presents the socio-demographic features and a comparison between all patients ("All" referring to the total patients in our database) and ADRD patients in Maryland. Compared to all patients, two features, age and total charge value, stand out as significantly different. This highlights that ADRD patients are generally older and that this disease is considerably more expensive to manage relative to other diseases.

Table 1 – Description of the Socio-Demographic Characteristics of ADRD and All Patients in Maryland

|  | ADRD | All |
|---|---|---|
| **Mean Age** | 79.08 | 46.6 |
| **Pct of Female** | 55.6 | 58.9 |
| **Pct White** | 60.8 | 49.5 |
| **Pct Black** | 32.6 | 31.1 |
| **Mean Amount of Charge $** | 29,808 | 23,238 |

**Figure 1** shows the distribution of hospitals in Maryland, as well as the distribution of the population and the percentage of Black people. As shown in the figure, there is a concentration of hospitals in central Maryland and more urban areas, while the distribution becomes sparse in the peripheral regions of the map. Moreover, A greater population of Black people and low-income communities lives in these areas with a higher number of hospitals. Furthermore, **Table 2** indicates that population size and the percentage of people living below the poverty line have a positive and significant relationship to ZIP codes with hospitals (If a hospital is on the edge of two ZIP codes, both ZIP codes are considered to have a hospital.)

Table 2 - Description of the correlation between the presence of hospitals in ZIP codes and population size, percentage of Black residents, and percentage of people living below the poverty line

| Relationship to ZIP code Containing a hospital | Correlation Value | P - Value |
|---|---|---|
| Population | 0.44 | ~ 0 |
| Percentage of Black people | 0.20 | ~ 0 |
| Percentage of people below poverty | 0.19 | ~ 0 |

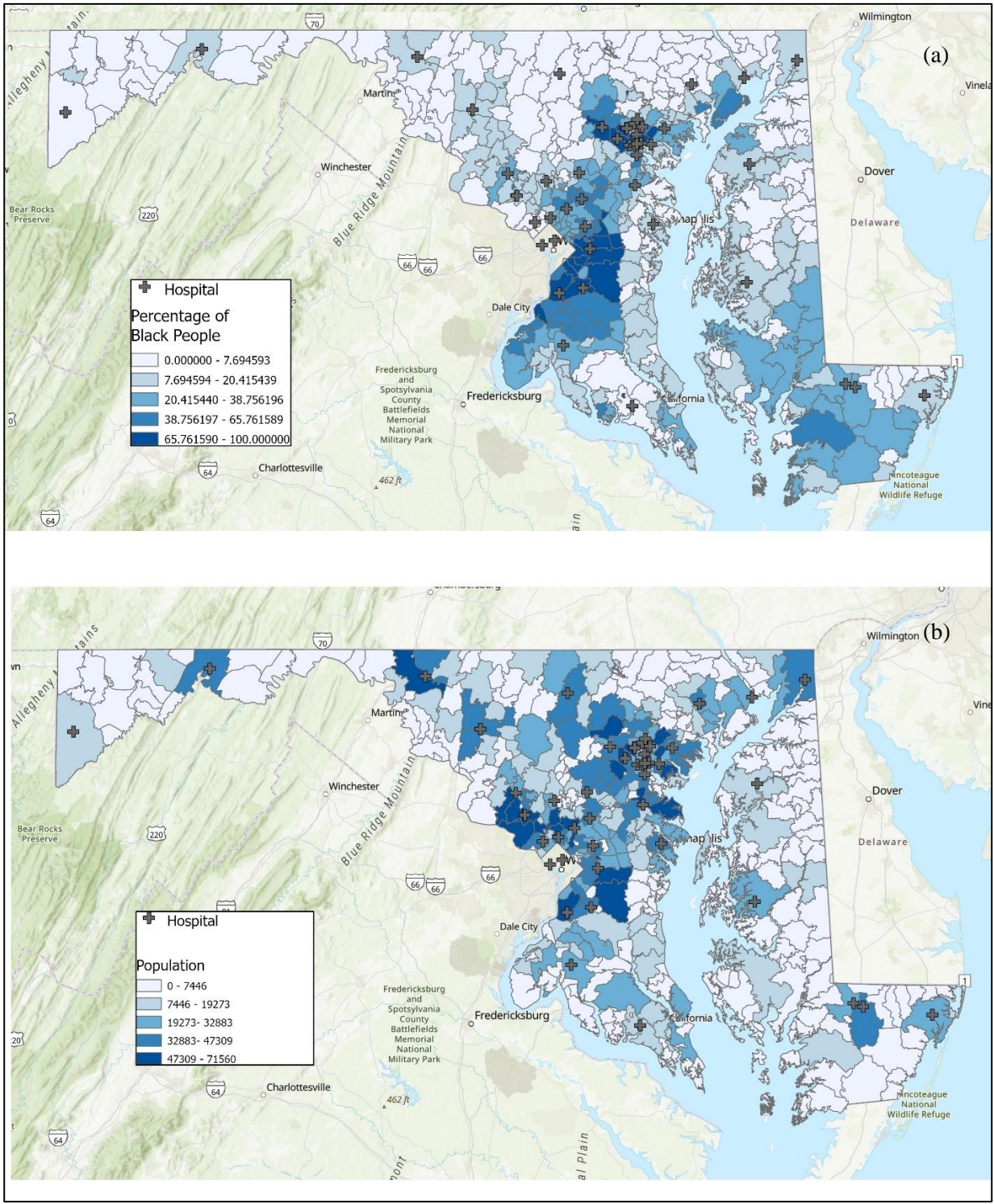

**Figure 1** - Distribution of (a) Black people and (b) population in Maryland

We calculated accessibility to hospitals from each ZIP code (**Figure 2**). The Gini index of 0.53 suggests that accessibility is unevenly distributed across the state. The Gini index is a measure of statistical dispersion that quantifies the degree of inequality in a distribution, ranging from 0 (perfect equality) to 1 (perfect inequality). A higher Gini index indicates greater inequality within the population. Not unexpectedly, we find that accessibility is significantly higher in urban areas compared to rural areas. It is also observed that, in addition to most hospitals being located near urban areas, healthcare facilities in rural areas are generally smaller and have fewer beds. Moreover, the Gini index reveals a highly unequal distribution of accessibility in rural areas compared to the more equal distribution observed in urban areas. (**Figure** *3*).

As mentioned, analyzing different variables across rural and urban areas is important for identifying inequalities between them. Social variables such as education level, age, health insurance coverage, economic indicators like poverty rate, and health-related variables such as the percentage of people with diabetes, which can be correlated with ADRD, are all relevant factors. Table 3 shows the Result of the t-test model for comparison between variables in rural and urban areas. It suggests that the average accessibility, percentage of people with higher education, and percentage of young people are significantly higher in urban areas compared to rural areas. Conversely, the average percentage of people over 50 years old and the percentage of people with health insurance are higher in rural areas. Interestingly, the percentage of people with diabetes, depression, hyperlipidemia, hypertension, and heart disease, all of which are identified as related conditions to ADRD (1), is significantly higher in rural areas.

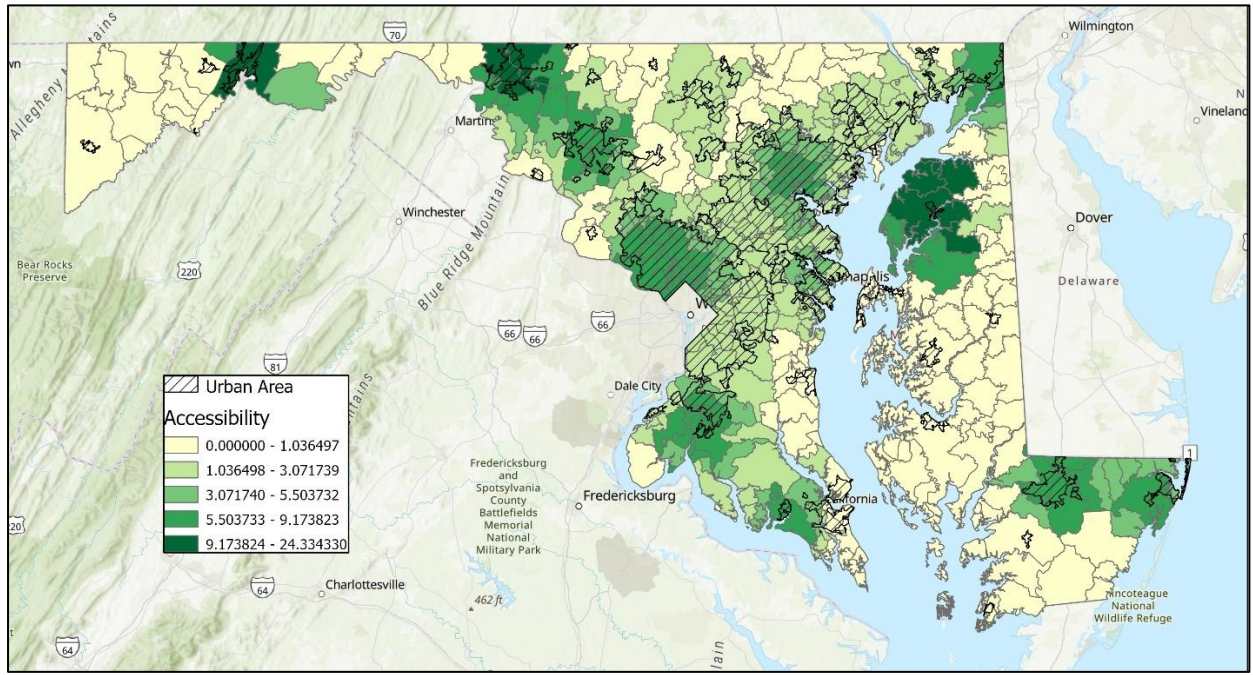

**Figure 2** - Distribution of accessibility to hospitals and urban areas in Maryland

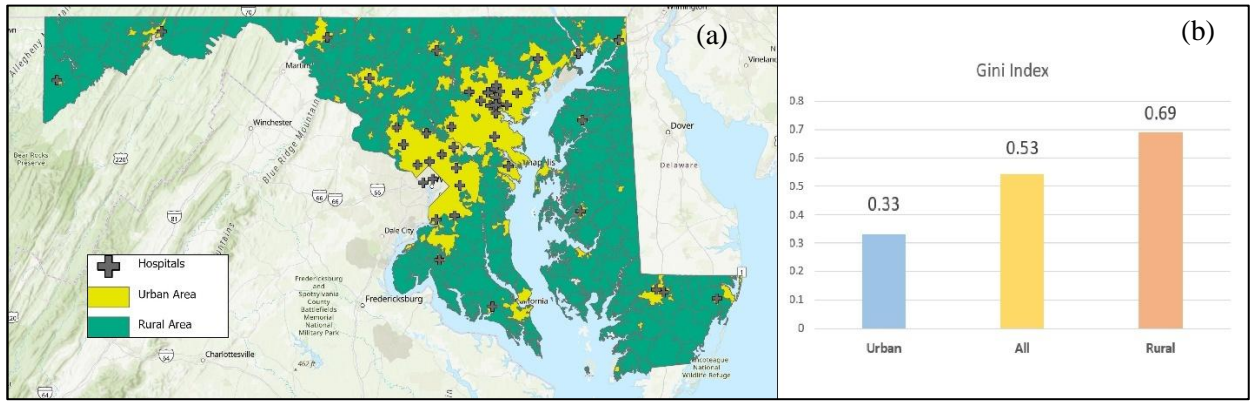

**Figure 3** - (a) Distribution of hospitals in rural and urban areas - (b) Gini index of accessibility in rural and urban areas

Table 3 - Results of the t-test for comparison between variables in rural and urban areas

|  | Mean value | | Variance Value | | | |
| --- | --- | --- | --- | --- | --- | --- |
| Variables | Urban | Rural | Urban | Rural | t test | P Value |
| Accessibility | 4.1035 | 2.1245 | 6.28 | 10.08 | -7.15 | 0.000 |
| People with high Education % | 15.5357 | 14.2861 | 35.61 | 81.83 | -1.71 | 0.086 |
| Young People % | 10.4080 | 8.1197 | 17.08 | 15.70 | -5.67 | 0.000 |
| People Over 50 % | 35.238 | 44.312 | 75.06 | 142.66 | 9.14 | 0 |
| People with Health Insurance % | 97.0943 | 98.9291 | 103.51 | 6.76 | 2.28 | 0.023 |
| Population | 25,820.75 | 6,894.35 | 28,212 | 9,831 | -12.74 | 0.000 |
| Below Poverty Rate % | 8.6745 | 7.2397 | 43.41 | 45.83 | -2.17 | 0.030 |
| People with Diabetes % | 0.0127 | 0.0143 | 0.0001 | 0.0001 | 1.78 | 0.074 |
| People with Depression % | 0.0087 | 0.0098 | 0.0000 | 0.0000 | 2.0306 | 0.043 |
| People with Hyperlipidemia % | 0.0203 | 0.0244 | 0.0001 | 0.0002 | 3.0790 | 0.002 |
| People with Hypertension % | 0.0275 | 0.0318 | 0.0002 | 0.0003 | 2.6410 | 0.008 |
| People with Heart Disease % | 0.0188 | 0.0218 | 0.0001 | 0.0002 | 2.5030 | 0.012 |

**Figure *4*-**A & Figure 4-B show the distribution of hospitals, the number of diagnosed ADRD patients, and the rate of ADRD patients to population of each ZIP code across Maryland. Both the number and proportion of ADRD patients are higher in hospital-rich ZIP codes. This may be related to the availability of better healthcare facilities, which can lead to more diagnoses of these patients.

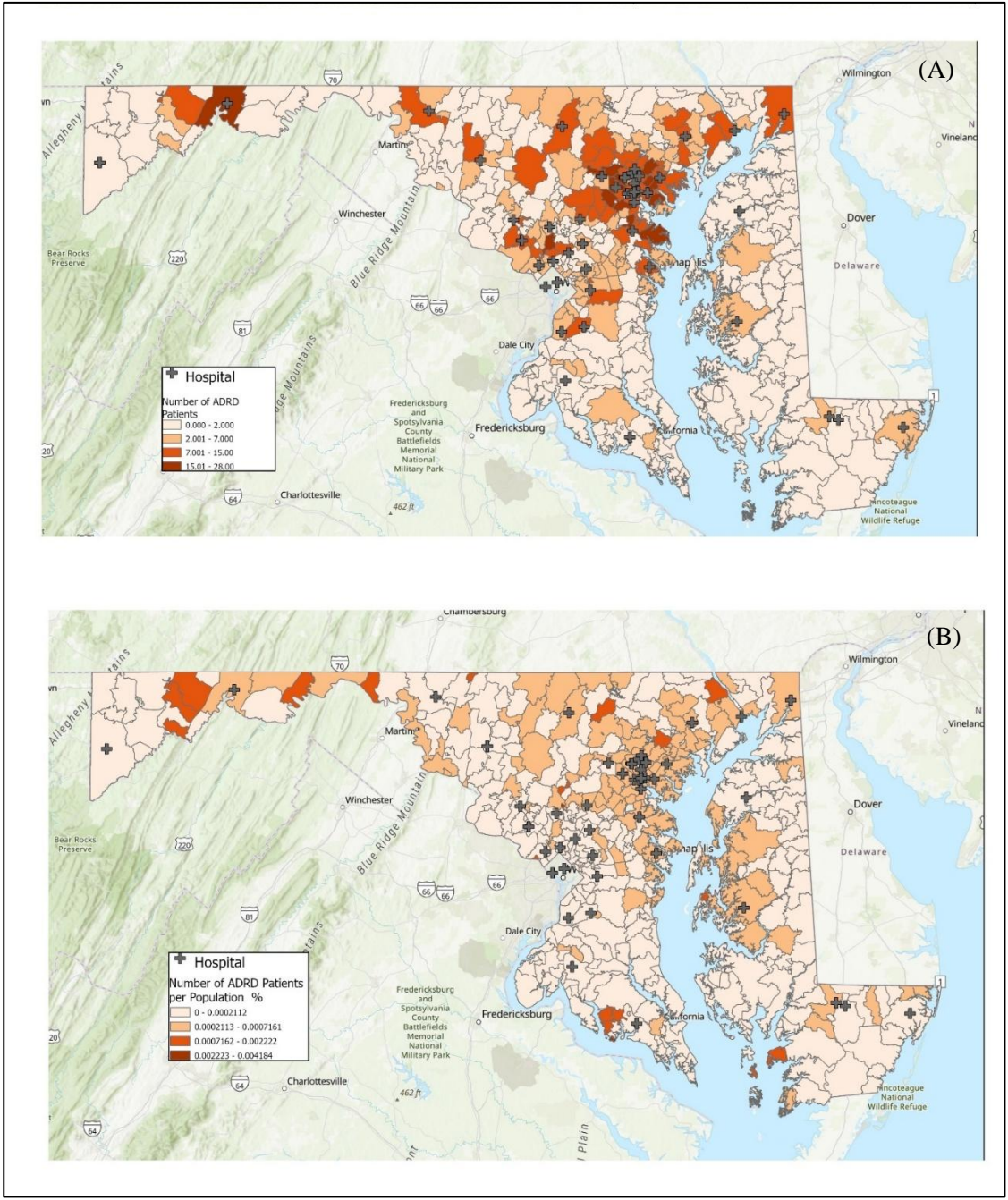

**Figure 4** - Distribution of (A) ratio of ADRD patients per population (B) Number of ADRD patients in Maryland

This also raises the question as to whether lows rate of diagnosed ADRD patients in low healthcare areas corresponds to low ADRD mortality rates. We looked at whether mortality rates due to ADRD were higher or lower in areas with less availability of hospitals (**Figure 5**). The eastern and southern regions of Maryland, which have fewer recorded ADRD patients, exhibit higher ADRD mortality rates, suggesting that these are underserved areas. In contrast, central Maryland, with a higher concentration of hospitals, reports greater numbers of ADRD patients but lower mortality rates due to ADRD. This might indicate several things, including better or quicker treatment access.

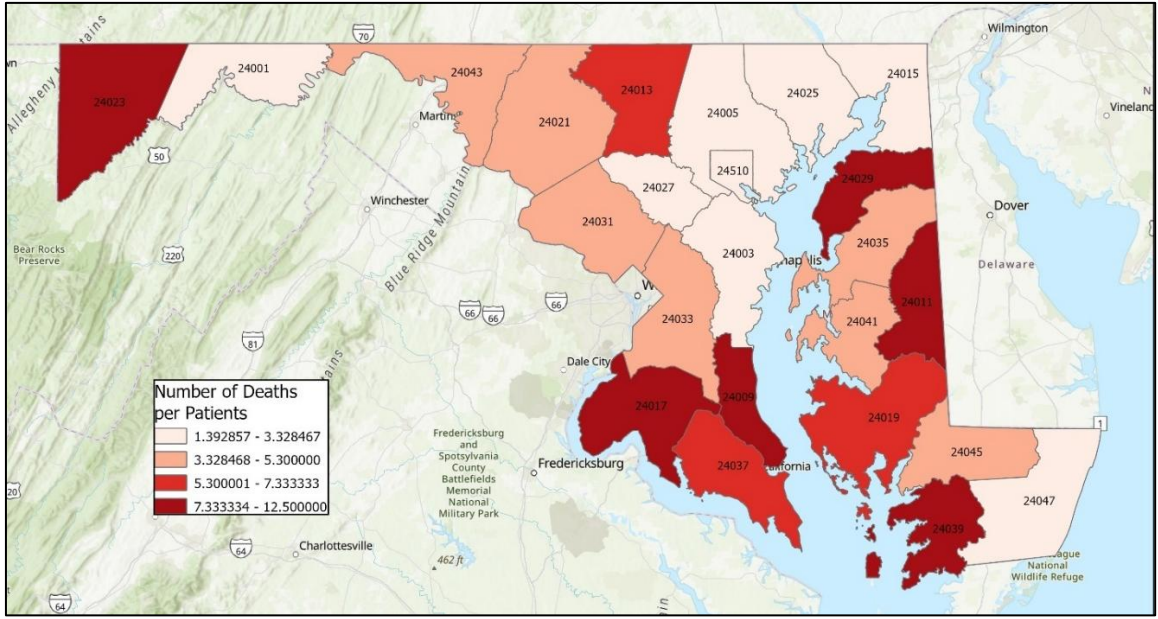

**Figure 5** - Distribution of the ratio of the number of deaths due to ADRD per number of ADRD patients at the county level in Maryland – Labeled with county number

This analysis focuses on the state of Maryland in 2023; however, the observed pattern is not limited to a single year. To gain a more in-depth understanding, the same methodology was applied to the years 2018 through 2022. We calculated the average number of deaths due to ADRD and the number of ADRD patients during this period. The figure shows the number of deaths per patient, revealing a similar trend across counties, with higher mortality rates due to ADRD observed in non-central areas of Maryland.

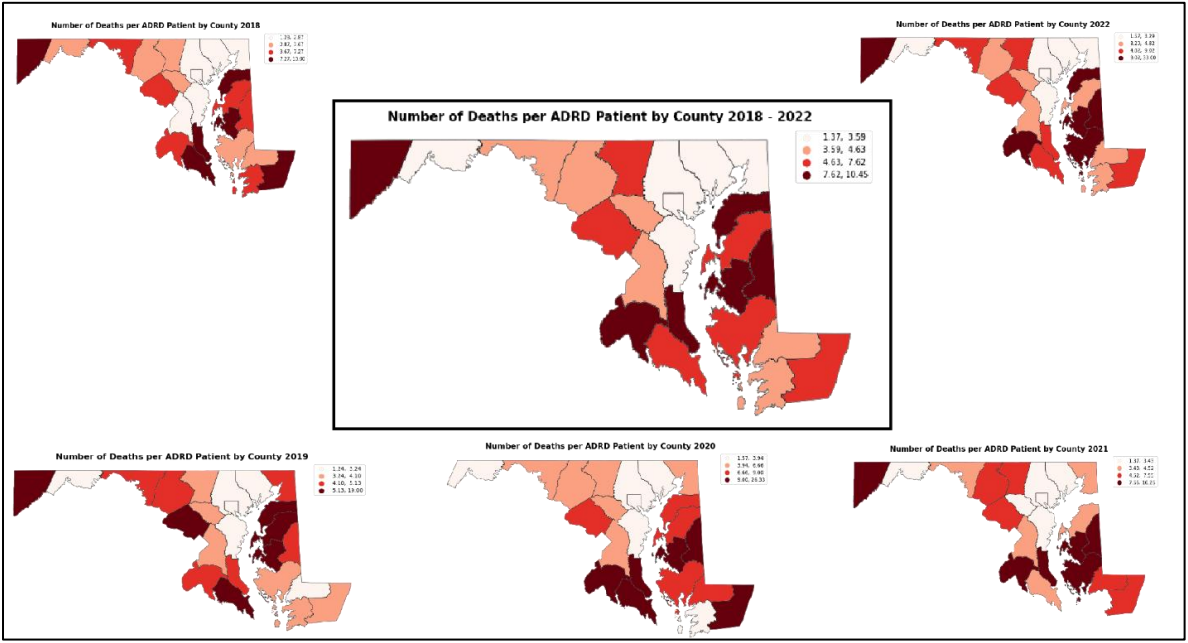

**Figure 6** - Distribution of the ratio of deaths due to ADRD per number of ADRD patients at the county level in Maryland for each year from 2018 to 2022, and the average ratio for 2018–2022.

**Figure 7** displays mortality rates due to ADRD and ADRD diagnosis rates at the county level across Maryland. Some counties exhibit mortality rates that are substantially higher than the average, indicating potentially underserved areas.

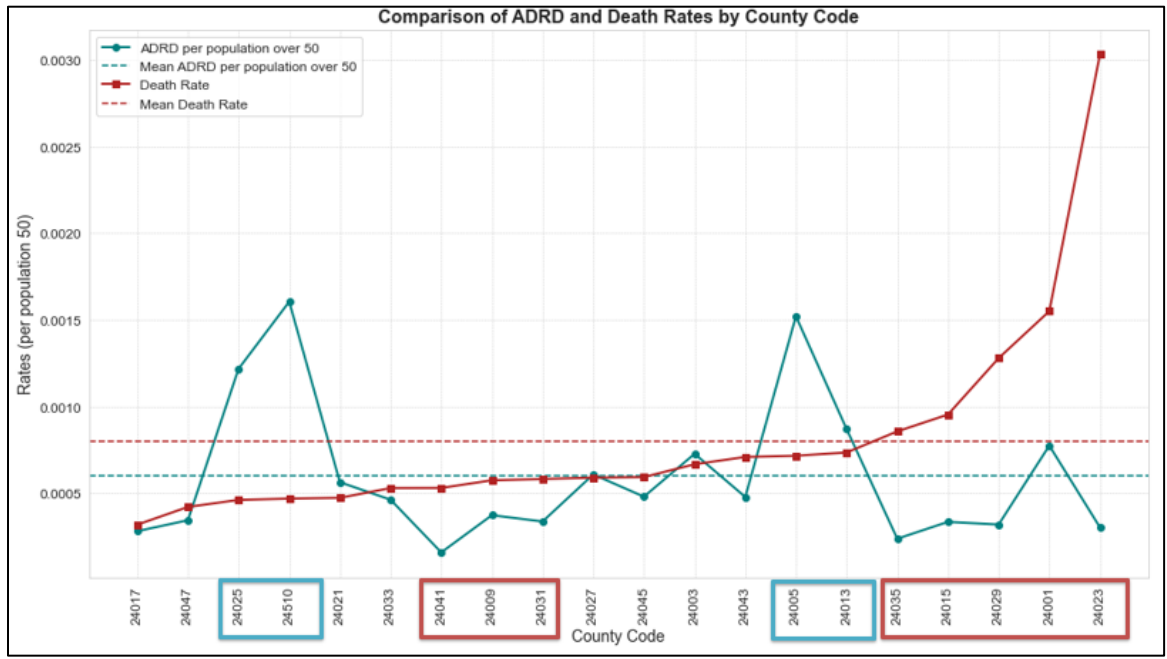

**Figure 7** - Comparison of the ratio of ADRD patients per population aged 50 and over with the mortality rate due to ADRD per population aged 50 and over at the county level

In the final step, we used the Getis-Ord Gi* model for hot spot analysis to identify hot and cold spots of two variables, accessibility and the number of deaths due to ADRD per number of ADRD patients, across Maryland (Figure 8). By calculating the Getis-Ord Gi* statistic, the resulting z-scores and p-values indicate where features with high or low values are spatially clustered. This method evaluates each feature in the context of its neighboring features. A feature with a high value may be interesting, but it is not necessarily a statistically significant hot spot. For a feature to be considered a statistically significant hot spot, it must have a high value and be surrounded by other features with similarly high values. Conversely, a statistically significant cold spot has a low value and is surrounded by other features with low values. (50) (51)

Figure 8 - A shows that in the eastern and southern regions of Maryland, there are statistically significant hot spots, where areas with high death (due to ADRD) rates among diagnosed ADRD patients are surrounded by neighbors with similarly high rates. On the other hand, the central regions of Maryland, mainly urban areas with better healthcare facilities, exhibit cold spots, indicating significantly lower death rates among diagnosed ADRD patients, surrounded by neighbors with similarly low rates. Moreover, the Gi* results for accessibility at the ZIP code level also confirm significant hot spots in central Maryland and cold spots in the eastern region of the state. (Figure 8 - B)

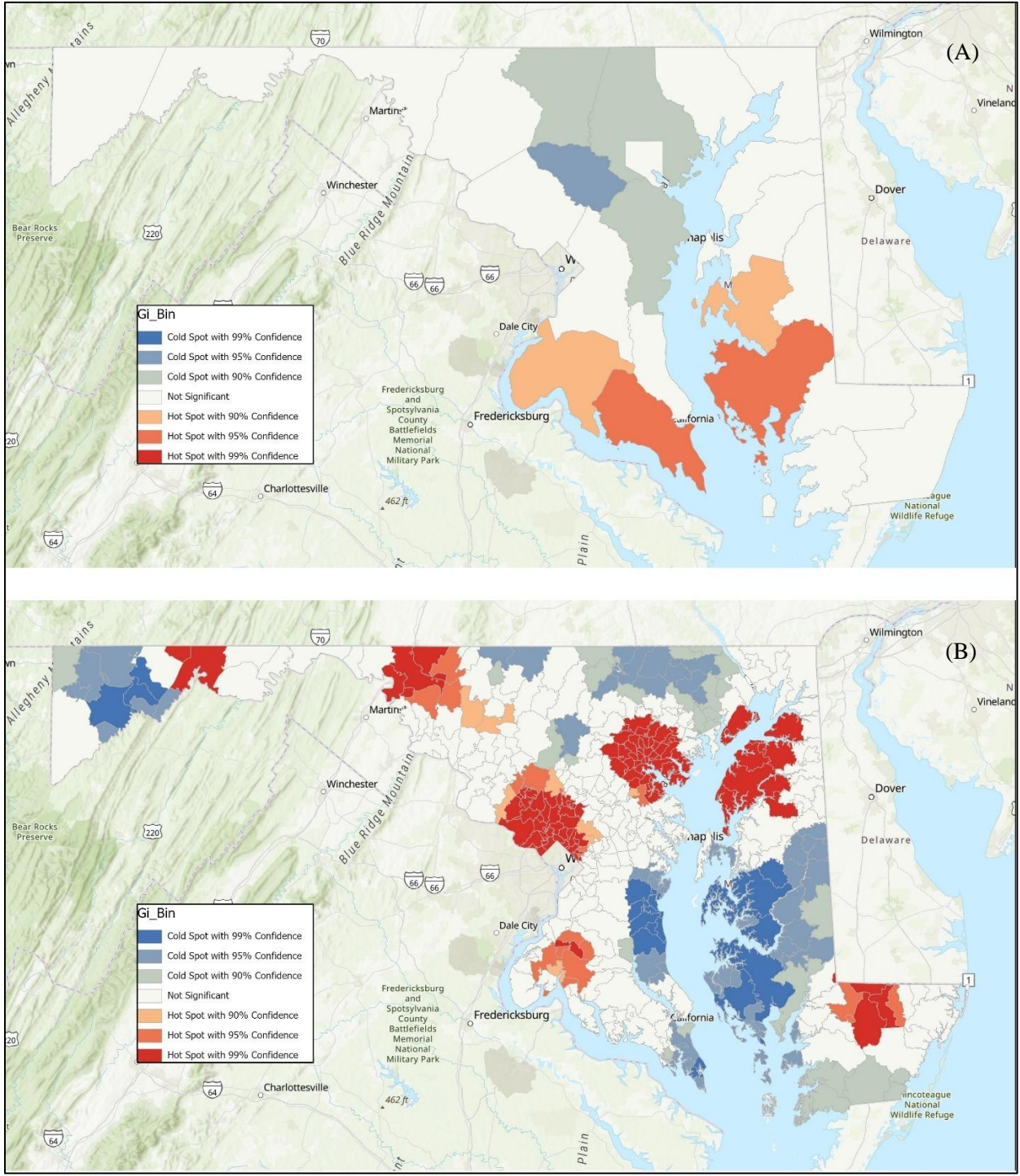

**Figure 8** - Getis-Ord Gi statistic for (A) the number of deaths due to ADRD per number of ADRD patients at the county level, and (B) accessibility at the ZIP code level in Maryland.

An important component of many geospatial analyses is the comparison of two variables across a study area to determine whether they are related and how that relationship manifests. Traditionally, such questions have been addressed through either careful cartographic comparison or linear regression analysis. However, cartographic comparison can be subjective, and regression analysis is limited to detecting only simple relationships. The Local Bivariate Relationships method quantifies the relationship between two variables on the same map by determining whether the values of one variable depend on or are influenced by the values of another, and whether those relationships vary across geographic space. (51)

Figure 9–A shows the local bivariate relationship results between the rate of ADRD patients and accessibility. In the eastern, southern, and far eastern regions (blue areas), there is a statistically significant relationship between lower accessibility and lower rates of diagnosed ADRD patients. Conversely, the red areas indicate regions where higher rates of diagnosed ADRD patients are significantly associated with higher accessibility. **Figure *9*-B** shows that in the eastern region (orange zones), the percentage of people below the poverty rate has a negative relationship with accessibility; areas with a higher percentage of people below the poverty rate tend to have lower accessibility to healthcare facilities. In contrast, in central Maryland (light blue zones), there is a significant relationship between lower poverty rates and higher accessibility. In terms of health risk, the Health Risk Index was calculated using PCA, capturing 75% of the total variance. After applying the Local Bivariate Relationships method, it was found that, once again, in the eastern region of Maryland, there are statistically significant zones with both high poverty rates and higher Health Risk Index values, indicating a higher prevalence of diseases related to ADRD. Conversely, in the central region of Maryland, areas with lower poverty rates are associated with lower Health Risk Index values (**Figure *9*-C**).

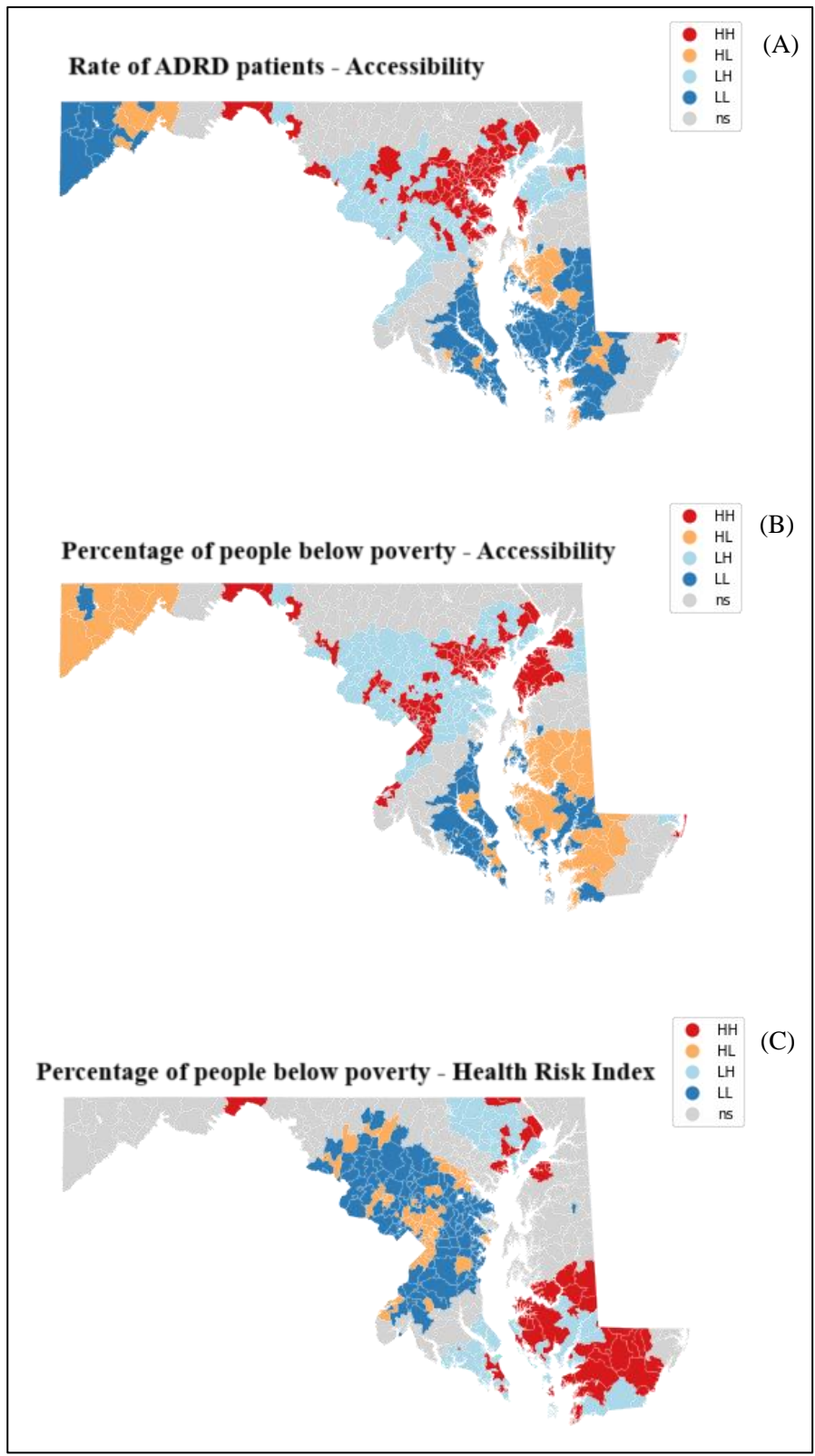

**Figure 9** - Local bivariate relationship results for (A) the rate of ADRD patients and accessibility, (B) the percentage of people below the poverty rate and accessibility, and (C) the percentage of people below the poverty rate and the health risk index

# Discussion

The findings of this study help address the question, "Are we really taking care of all patients?" The results reveal that, despite having more hospitals in areas with higher populations of socially and economically vulnerable individuals, the distribution of hospital accessibility creates significant inequality within the community, particularly in rural areas. A deeper analysis highlights the existence of underserved areas where mortality rates due to ADRD are much higher than the rates of diagnosed ADRD. It is particularly important to note that counties in the eastern and southern regions of Maryland experience high mortality rates due to ADRD but report low rates of diagnosed ADRD. This raises a critical question: Does the lack of accessibility prevent the identification and diagnosis of ADRD patients in these areas? The high mortality rates due to ADRD suggest that patients in these regions are indeed suffering from the disease, even if they remain undiagnosed. This conclusion underscores the need for increased attention and investment in healthcare facilities in these underserved counties to ensure equitable access to care and improve health outcomes.

Based on the description of ADRD and other patients in Maryland (from the HCUP dataset), it can be observed that, as expected (52), the average age of ADRD patients is close to 80 years. This emphasizes the critical importance of ensuring convenient accessibility to healthcare for these patients, who are among the most vulnerable. Furthermore, the average total charge for ADRD patients is approximately 1.4 times higher than the average total charge for other patients. This finding is consistent with previous research (53). This highlights the need to better understand the factors influencing ADRD and to develop more effective strategies for addressing this disease earlier. Doing so could help prevent additional costs and challenges for both patients and healthcare facilities.

The distribution of hospitals across Maryland reveals a significant concentration in urban areas, with most of the hospitals located in designated urban regions. This finding is consistent with national trends, where urban areas typically have a higher density of healthcare facilities compared to rural areas (54). This uneven distribution contributes to disparities in healthcare accessibility, particularly in rural areas. The results indicate that the Gini index for accessibility in rural areas is approximately 0.7, highlighting the unequal distribution of healthcare access, which also confirms previous studies (55). Measuring and evaluating equitable access to healthcare services across the population is essential for enhancing public health and strengthening the effectiveness of national health systems. Rural–urban disparities in access to healthcare are prevalent in many countries. Numerous recent studies have explored and confirmed the existence of inequalities between rural and urban areas, highlighting the significance of this issue. (56) (57) (58) (59). Table 3 presents the t-test results for various variables in both rural and urban areas. It is important to note that, despite older adults being the most vulnerable to ADRD, urban areas with greater hospital accessibility tend to have a higher proportion of younger individuals and a lower proportion of older individuals compared to rural areas. Moreover, in rural areas, the average percentage of people with diseases such as diabetes, obesity, asthma, depression, hyperlipidemia, hypertension, and heart disease, which can be correlated with ADRD (1), is significantly higher than in urban areas. This underscores the critical need to focus more attention on improving healthcare accessibility in rural areas.

The hot spot analysis highlights a critical issue: how low accessibility can contribute to a higher mortality rate due to ADRD. At first glance, the relationship between accessibility and the rate of ADRD patients appears to suggest that regions with lower accessibility have fewer diagnosed ADRD patients, while regions with higher accessibility have higher rates of diagnosis. However, a deeper analysis using the Gi* statistic reveals that areas with lower accessibility actually experience significantly higher death rates due to ADRD. This finding underscores the urgent need to address healthcare inequality and better serve these underserved regions.

There are some limitations to this study. First, the analysis is limited to the state of Maryland. Expanding the sample size to include multiple states or even a nationwide analysis would provide a broader perspective and yield more robust results by better capturing the relationships between variables. Second, conducting a more detailed analysis of how individual patients respond to the disease could provide valuable insights. Third, the analysis and accessibility calculations in this study are based on previous research and assume a constant distance threshold. Performing the analysis with varying thresholds and conducting sensitivity analyses could enhance the understanding of accessibility-related disparities. Finally, having more detailed information about hospitals could improve the accuracy and reliability of the results. In this study, hospital data were limited to the number of beds, which was used to calculate accessibility. However, incorporating additional details, such as the presence of specialized ADRD units or the number of ADRD experts in each hospital, could significantly enhance the research.

## Conclusion

In conclusion, the findings indicate that although there are more hospitals in areas with higher concentrations of socially and economically vulnerable populations, the unequal distribution of hospital accessibility leads to significant disparities, especially in rural regions. The analysis reveals underserved areas where mortality rates from ADRD are notably higher than the rates of diagnosed cases. Notably, counties in the eastern and southern parts of Maryland show high mortality rates due to ADRD despite having low diagnosis rates. These observations imply that patients in these areas are likely suffering from ADRD without being diagnosed. Therefore, there is a need for enhanced focus and investment in healthcare facilities within these underserved counties to promote fair access to care and improve overall health outcomes. The hot spot analysis confirms that there are regions in the eastern and southern parts of Maryland with significantly high death rates per diagnosed patient, surrounded by neighboring areas with similarly high values. In contrast, cold spots are observed in the central region of Maryland. The local bivariate analysis further confirms that in the eastern region, a higher percentage of people below the poverty rate is significantly associated with a higher Health Risk Index and lower accessibility. This helps to better understand where to prioritize investment and identify areas with stronger associations with ADRD.

### Author Contributions

The authors confirm contribution to the paper as follows: study conception and design: S. Saleh Namadi, J Chen, D. Niemeier; data collection: S. Saleh Namadi, J. Chen; analysis and interpretation of results: S. Saleh Namadi, D. Niemeier; draft manuscript preparation: S. Saleh Namadi, J. Chen, D. Niemeier. All authors reviewed the results and approved the final version of the manuscript.

# References

1. Wang N, Maguire TK, Chen J. Preventable Emergency Department Visits of Patients with Alzheimer's Disease and Related Dementias During the COVID-19 Pandemic by Hospital-Based Health Information Exchange. Gerontol Geriatr Med. 2024 Jan 1;10:23337214241244984.

2. Nicholls SG, Al-Jaishi AA, Niznick H, Carroll K, Madani MT, Peak KD, et al. Health equity considerations in pragmatic trials in Alzheimer's and dementia disease: Results from a methodological review. Alzheimers Dement Diagn Assess Dis Monit. 2023;15(1):e12392.

3. Dementia [Internet]. [cited 2025 Apr 1]. Available from: https://www.who.int/news-room/fact-sheets/detail/dementia

4. Quiñones AR, Kaye J, Allore HG, Botoseneanu A, Thielke SM. An Agenda for Addressing Multimorbidity and Racial and Ethnic Disparities in Alzheimer's Disease and Related Dementia. Am J Alzheimers Dis Dementias®. 2020 Jan 1;35:1533317520960874.

5. ISPOR | International Society For Pharmacoeconomics and Outcomes Research [Internet]. [cited 2025 Apr 1]. Value Assessment in Alzheimer's Disease: A Focus on Equity. Available from: https://www.ispor.org/publications/journals/value-outcomes-spotlight/vos-archives/issue/view/valuing-future-alzheimers-disease-treatments-the-need-for-a-holistic-approach/value-assessment-in-alzheimers-disease-a-focus-on-equity

6. Chin AL, Negash S, Hamilton R. Diversity and Disparity in Dementia: The Impact of Ethnoracial Differences in Alzheimer Disease. Alzheimer Dis Assoc Disord. 2011 Sep;25(3):187.

7. Institute of Health Equity [Internet]. [cited 2025 Apr 1]. Inequalities in Mental Health, Cognitive Impairment and Dementia Among Older People. Available from: https://www.instituteofhealthequity.org/resources-reports/inequalities-in-mental-health-cognitive-impairment-and-dementia-among-older-people

8. Mielke MM. Sex and Gender Differences in Alzheimer's Disease Dementia. Psychiatr Times. 2018 Nov;35(11):14–7.

9. 2021 Alzheimer's disease facts and figures - 2021 - Alzheimer's & Dementia - Wiley Online Library [Internet]. [cited 2025 Apr 1]. Available from: https://alz-journals.onlinelibrary.wiley.com/doi/10.1002/alz.12328

10. Chen J, Spencer MRT, Buchongo P, Wang MQ. Hospital-based Health Information Technology Infrastructure: Evidence of Reduced Medicare Payments and Racial Disparities Among Patients With ADRD. Med Care. 2023 Jan;61(1):27.

11. Gilligan AM, Malone DC, Warholak TL, Armstrong EP. Health Disparities in Cost of Care in Patients With Alzheimer's Disease: An Analysis Across 4 State Medicaid Populations. Am J Alzheimers Dis Dementias®. 2013 Feb 1;28(1):84–92.

12. Nicholl J, West J, Goodacre S, Turner J. The relationship between distance to hospital and patient mortality in emergencies: an observational study. Emerg Med J. 2007 Sep 1;24(9):665–8.

13. Mobaderi T, Kazemnejad A, Salehi M. Exploring the impacts of risk factors on mortality patterns of global Alzheimer's disease and related dementias from 1990 to 2021. Sci Rep. 2024 Jul 6;14(1):15583.

14. Agüero-Torres H, Fratiglioni L, Guo Z, Viitanen M, Winblad B. Mortality from Dementia in Advanced Age: A 5-Year Follow-Up Study of Incident Dementia Cases. J Clin Epidemiol. 1999 Aug 1;52(8):737–43.

15. Beeri MS, Goldbourt U. Late-Life Dementia Predicts Mortality Beyond Established Midlife Risk Factors. Am J Geriatr Psychiatry. 2011 Jan 1;19(1):79–87.

16. Risk of Death Among Persons with Alzheimer's Disease: A National Register-Based Nested Case-Control Study - Eija Lönnroos, Pentti Kyyrönen, J. Simon Bell, Tischa J.M. van der Cammen, Sirpa Hartikainen, 2013 [Internet]. [cited 2025 Apr 1]. Available from: https://journals.sagepub.com/doi/abs/10.3233/JAD-2012-120808

17. Livingston G, Huntley J, Sommerlad A, Ames D, Ballard C, Banerjee S, et al. Dementia prevention, intervention, and care: 2020 report of the *Lancet* Commission. The Lancet. 2020 Aug 8;396(10248):413–46.

18. Gu Z, Luo X, Tang M, Liu X. Does the edge effect impact the healthcare equity? An examination of the equity in hospitals accessibility in the edge city in multi-scale. J Transp Geogr. 2023 Jan 1;106:103513.

19. Scopus - Document details - Accessibility to primary health care in Belgium: An evaluation of policies awarding financial assistance in shortage areas [Internet]. [cited 2025 Apr 1]. Available from: https://www.scopus.com/record/display.uri?eid=2-s2.0-84882299910&origin=inward

20. Widener MJ, Hatzopoulou M. Contextualizing research on transportation and health: A systems perspective. J Transp Health. 2016 Sep 1;3(3):232–9.

21. Guagliardo MF. Spatial accessibility of primary care: concepts, methods and challenges. Int J Health Geogr. 2004 Feb 26;3(1):3.

22. Geurs KT, van Wee B. Accessibility evaluation of land-use and transport strategies: review and research directions. J Transp Geogr. 2004 Jun 1;12(2):127–40.

23. Mehditabrizi A, Tahmasbi B, Namadi SS, Cirillo C. En route and home proximity in EV charging accessibility: a spatial equity analysis. Transp Res Part Transp Environ. 2025 Sep 1;146:104910.

24. Neutens T. Accessibility, equity and health care: review and research directions for transport geographers. J Transp Geogr. 2015 Feb 1;43:14–27.

25. Luo J, Chen G, Li C, Xia B, Sun X, Chen S. Use of an E2SFCA Method to Measure and Analyse Spatial Accessibility to Medical Services for Elderly People in Wuhan, China. Int J Environ Res Public Health. 2018 Jul;15(7):1503.

26. Luo S, Jiang H, Yi D, Liu R, Qin J, Liu Y, et al. PM2SFCA: Spatial Access to Urban Parks, Based on Park Perceptions and Multi-Travel Modes. A Case Study in Beijing. ISPRS Int J Geo-Inf. 2022 Sep;11(9):488.

27. Luo W, Wang F. Measures of Spatial Accessibility to Health Care in a GIS Environment: Synthesis and a Case Study in the Chicago Region. Environ Plan B Plan Des. 2003 Dec 1;30(6):865–84.

28. Wang F. Measurement, Optimization, and Impact of Health Care Accessibility: A Methodological Review. Ann Assoc Am Geogr Assoc Am Geogr. 2012;102(5):1104–12.

29. Chen X, and Jia P. A comparative analysis of accessibility measures by the two-step floating catchment area (2SFCA) method. Int J Geogr Inf Sci. 2019 Sep 2;33(9):1739–58.

30. Zhang S, Song X, Zhou J. An equity and efficiency integrated grid-to-level 2SFCA approach: spatial accessibility of multilevel healthcare. Int J Equity Health. 2021 Oct 19;20(1):229.

31. Liu L, Zhao Y, Lyu H, Chen S, Tu Y, Huang S. Spatial Accessibility and Equity Evaluation of Medical Facilities Based on Improved 2SFCA: A Case Study in Xi'an, China. Int J Environ Res Public Health. 2023 Jan;20(3):2076.

32. Tao Z, Cheng Y, Liu J. Hierarchical two-step floating catchment area (2SFCA) method: measuring the spatial accessibility to hierarchical healthcare facilities in Shenzhen, China. Int J Equity Health. 2020 Sep 21;19(1):164.

33. Khashoggi BF, Murad A. Use of 2SFCA Method to Identify and Analyze Spatial Access Disparities to Healthcare in Jeddah, Saudi Arabia. Appl Sci. 2021 Jan;11(20):9537.

34. Shao Y, Luo W. Supply-demand adjusted two-steps floating catchment area (SDA-2SFCA) model for measuring spatial access to health care. Soc Sci Med. 2022 Mar 1;296:114727.

35. Kanuganti S, Sarkar AK, Singh AP. Quantifying Accessibility to Health Care Using Two-step Floating Catchment Area Method (2SFCA): A Case Study in Rajasthan. Transp Res Procedia. 2016 Jan 1;17:391–9.

36. Geospatial analysis of dementia mortality and its associations with demographic and socioeconomic factors [Internet]. [cited 2025 Apr 1]. Available from: https://journals.sagepub.com/doi/epub/10.3233/ADR-240107

37. Saleh Namadi S, Tahmasbi B, Mehditabrizi A, Darzi A, Niemeier D. Using Geographically Weighted Models to Explore Temporal and Spatial Varying Impacts on Commute Trip Change Resulting from COVID-19. Transp Res Rec. 2024 Dec 1;2678(12):687–701.

38. Firouraghi N, Kiani B, Jafari HT, Learnihan V, Salinas-Perez JA, Raeesi A, et al. The role of geographic information system and global positioning system in dementia care and research: a scoping review. Int J Health Geogr. 2022 Aug 4;21(1):8.

39. HCUP-US SID Overview [Internet]. [cited 2025 Jul 23]. Available from: https://hcup-us.ahrq.gov/sidoverview.jsp

40. American Hospital Association Homepage | Hospitals USA | AHA [Internet]. 2025 [cited 2025 Jul 23]. Available from: https://www.aha.org/

41. CDC WONDER [Internet]. [cited 2025 Jul 23]. Available from: https://wonder.cdc.gov/

42. Bureau USC. American community survey (ACS). Census Gov Retrieved Novemb. 2021;1:2021.

43. Kramarow EA, Tejada-Vera B. Dementia Mortality in the United States, 2000-2017. Natl Vital Stat Rep Cent Dis Control Prev Natl Cent Health Stat Natl Vital Stat Syst. 2019 Mar;68(2):1–29.

44. Bauer J, Groneberg DA. Measuring Spatial Accessibility of Health Care Providers – Introduction of a Variable Distance Decay Function within the Floating Catchment Area (FCA) Method. PLOS ONE. 2016 Jul 8;11(7):e0159148.

45. Cheng G, Zeng X, Duan L, Lu X, Sun H, Jiang T, et al. Spatial difference analysis for accessibility to high level hospitals based on travel time in Shenzhen, China. Habitat Int. 2016 Apr 1;53:485–94.

46. Kwan MP. Space-Time and Integral Measures of Individual Accessibility: A Comparative Analysis Using a Point-based Framework. Geogr Anal. 1998;30(3):191–216.

47. Fernández-Cardero Á, Sierra-Cinos JL, López-Jiménez A, Beltrán B, Cuadrado C, García-Conesa MT, et al. Characterizing Factors Associated with Excess Body Weight: A Descriptive Study Using Principal Component Analysis in a Population with Overweight and Obesity. Nutrients. 2024 Jan;16(8):1143.

48. Agarwal S, Jacobs DR, Vaidya D, Sibley CT, Jorgensen NW, Rotter JI, et al. Metabolic Syndrome Derived from Principal Component Analysis and Incident Cardiovascular Events: The Multi Ethnic Study of Atherosclerosis (MESA) and Health, Aging, and Body Composition (Health ABC). Cardiol Res Pract. 2012;2012:919425.

49. Jolliffe IT, Cadima J. Principal component analysis: a review and recent developments. Philos Transact A Math Phys Eng Sci. 2016 Apr 13;374(2065):20150202.

50. Getis A, Ord JK. The Analysis of Spatial Association by Use of Distance Statistics. Geogr Anal. 1992;24(3):189–206.

51. ArcGIS Pro geoprocessing tool reference—ArcGIS Pro | Documentation [Internet]. [cited 2025 Aug 4]. Available from: https://pro.arcgis.com/en/pro-app/latest/tool-reference/main/arcgis-pro-tool-reference.htm

52. Fishman E. Risk of Developing Dementia at Older Ages in the United States. Demography. 2017 Oct;54(5):1897–919.

53. Kamdar N, Syrjamaki J, Aikens JE, Mahmoudi E. Readmission Rates and Episode Costs for Alzheimer Disease and Related Dementias Across Hospitals in a Statewide Collaborative. JAMA Netw Open. 2023 Mar 16;6(3):e232109.

54. Coombs NC, Campbell DG, Caringi J. A qualitative study of rural healthcare providers' views of social, cultural, and programmatic barriers to healthcare access. BMC Health Serv Res. 2022 Apr 2;22:438.

55. Xu R, Yue W, Wei F, Yang G, Chen Y, Pan K. Inequality of public facilities between urban and rural areas and its driving factors in ten cities of China. Sci Rep. 2022 Aug 2;12(1):13244.

56. Song Y, Tan Y, Song Y, Wu P, Cheng JCP, Kim MJ, et al. Spatial and temporal variations of spatial population accessibility to public hospitals: a case study of rural–urban comparison. GIScience Remote Sens. 2018 Sep 3;55(5):718–44.

57. Haggerty JL, Roberge D, Lévesque JF, Gauthier J, Loignon C. An exploration of rural–urban differences in healthcare-seeking trajectories: Implications for measures of accessibility. Health Place. 2014 Jul 1;28:92–8.

58. Du X, Du Y, Zhang Y, Zhu Y, Yang Y. Urban and rural disparities in general hospital accessibility within a Chinese metropolis. Sci Rep. 2024 Oct 7;14(1):23359.

59. McCarthy S, Moore D, Smedley WA, Crowley BM, Stephens SW, Griffin RL, et al. Impact of Rural Hospital Closures on Health-Care Access. J Surg Res. 2021 Feb 1;258:170–8.